# 基於後量子密碼學的 X.509 資安憑證

# X.509 Information Security Certification Based on Post-Quantum Cryptography

Abel C. H. Chen
Chunghwa Telecom Laboratories
ORCID: 0000-0003-3628-3033

## 摘要

近年來，隨著量子計算的發展，目前公開金鑰基礎建設(Public Key Infrastructure, PKI)系統中的主流非對稱式密碼學方法逐漸受到威脅。有鑑於此，本研究探討基於後量子密碼學(Post-Quantum Cryptography, PQC)的 X.509 資安憑證，並且討論和實現可行的解決方案。本研究比較主流非對稱式密碼學方法(包括 RSA 和橢圓曲線數位簽章演算法(Elliptic Curve Digital Signature Algorithm, ECDSA))與標準後量子密碼方法(包括 Falcon、Dilithium、SPHINCS+)，分別比較產製憑證、產製簽章、以及驗證簽章的效率。最後，提出基於後量子密碼學的 X.509 資安憑證解決方案建議。

**關鍵詞**：後量子密碼學、X.509 資安憑證、公開金鑰基礎建設、非對稱式密碼學。

## Abstract

In recent years, with the advancement of quantum computing, mainstream asymmetric cryptographic methods in the current Public Key Infrastructure (PKI) systems are gradually being threatened. Therefore, this study explores X.509 security certificates based on Post-Quantum Cryptography (PQC) and discusses implemented solutions. This study compares mainstream asymmetric cryptographic methods (including RSA and Elliptic Curve Digital Signature Algorithm (ECDSA)) with standard PQC methods (including Falcon, Dilithium, SPHINCS+), comparing the efficiency of certificate generation, signature generation, and signature verification. Finally, recommendations for a solution based on PQC for X.509 security certificates are proposed.

**Keywords:** Post-Quantum Cryptography, X.509 Security Certificate, Public Key Infrastructure, Asymmetric Cryptographic.

## 一、前言

公開金鑰基礎建設(Public Key Infrastructure, PKI)系統[1]通過非對稱式密碼學方法(如：RSA [2]和橢圓曲線數位簽章演算法(Elliptic Curve Digital Signature Algorithm, ECDSA)[3]等)建立可信任的基礎設施，並且使用憑證作為信任的基礎。在 PKI 系統中，存在根憑證中心(Root Certificate

Authority, RCA)，可以自簽其憑證。PKI 系統中的全部終端設備(End Entity, EE)都信任根憑證中心發行的憑證。根憑證中心可以為 PKI 系統中的終端設備發行憑證。這些憑證包含憑證所有者的公鑰，並且由根憑證中心簽署，把簽章放在終端設備的憑證中。因此，當其他終端設備收到憑證時，它們可以使用根憑證中心的公鑰來驗證這些憑證中的簽章，作為信任的基礎。

除了包含公鑰和簽章等信息外，憑證還需要存放數位簽章演算法(Digital Signature Algorithm, DSA)、所有者信息、權限、到期日期和其他相關信息的詳細信息。因此，有必要制定標準規範來定義憑證的格式，而 X.509 憑證格式[4]即是被提出且廣泛使用於此目的。此外，由於存在各種數位簽章演算法及其不同的參數組合，所以需要物件識別碼(Object Identifier, OID)[5]來描述數位簽章演算法及其相關參數，以提供給其他終端設備根據相關資訊來自動化驗證憑證。

儘管現行的 PKI 系統和 X.509 憑證提供了可信任的基礎設施，但它們是建立在 RSA 加密和橢圓曲線密碼學(Elliptic-Curve Cryptography, ECC)加密的基礎上。然而，量子計算技術已經逐漸對現行的 PKI 系統和非對稱式密碼學構成了威脅[6]。因此，美國國家標準暨技術研究院(National Institute of Standards and Technology, NIST)開始尋找後量子密碼學(Post-Quantum Cryptography, PQC)方法，以抵禦量子計算攻擊[7]。後量子密碼學方法主要包括基於格(Lattice-Based)、基於雜湊(Hash-Based)、基於編碼(Code-Based)和基於多變量(Multivariate-Based)的密碼學，每種方法都建立在不同的數學理論基礎上確保安全。然而，這些方法在產製金鑰時間、產製簽章時間和驗證簽章時間上可能存在差異。因此，本研究旨在探索各種後量子密碼學方法以構建 X.509 憑證並驗證系統效能，作為未來 PKI 系統開發人員參考。

本文分為四節。第二節概述各種密碼學方法，分為易受量子攻擊密碼學(Quantum-Vulnerable Cryptography, QVC)和後量子密碼學。第三節介紹基於後量子密碼學的 X.509 資安憑證概念，並且提供案例研究。第四節驗證基於後量子密碼學的 X.509 資安憑證產製憑證效率、產製簽章效率和驗證簽章效率。最後，第五節總結了本研究的貢獻，並且討論未來可行的研究方向。

## 二、密碼學方法

第1小節和第2小節介紹易受量子攻擊密碼學，包含 RSA 密碼學和橢圓曲線密碼學。第3小節~第6小節介紹後量子密碼學，包含基於格的密碼學、基於雜湊的密碼學、基於編碼的密碼學和基於多變量的密碼學。

### 1. 易受量子攻擊密碼學—RSA

在 RSA 密碼學中，首先選擇兩個質數 $a_1$ 和 $a_2$，並基於 $a_1$ 和 $a_2$ 生成一個模數(modulus) $n_r$，如公式(1)所示，並將 $a_1$ 和 $a_2$ 代入公式(2)來生成模數 $\phi_r$。然後，可以基於模數 $\phi_r$ 產製私鑰 $p_r$ 和公鑰 $P_r$，且私鑰和公鑰滿足公式(3)中定義的特性。之後，私鑰 $p_r$ 可以用於為明文雜湊值 $x$ 產製簽章 $y_r$，如公式(4)所示；並且可以使用公鑰 $P_r$ 驗證簽章，如公式(5)所示[8]。然而，Shor 提出可以在多項式時間複雜度內解決質因數分解問題的量子演算法[6]，所以 RSA 密碼學方法是易受量子攻擊密碼學方法之一。

$$n_r = a_1 \times a_2 \tag{1}$$

$$\varphi_r = (a_1 - 1) \times (a_2 - 1) \tag{2}$$

$$p_r P_r \equiv 1 (\bmod \varphi_r) \tag{3}$$
$$y_r = x^{p_r} \tag{4}$$
$$x = y^{P_r} \tag{5}$$

## 2. 易受量子攻擊密碼學—橢圓曲線密碼學

在橢圓曲線密碼學中，首先選擇橢圓曲線參數(如：NIST P-256)；其中，包括基點坐標 $G$ 和橢圓曲線的階 $n_e$。產製隨機整數 $p_e$ 作為私鑰，並根據公式(6)產製公鑰 $P_e$ [9]。橢圓曲線數位簽章演算法可以產製另一個隨機整數 $r_e$ 來產製橢圓曲線點 $R_e$ (座標為$(x_r, y_r)$)，如公式(7)所示；然後，搭配私鑰 $p_e$ 根據公式(8)為明文雜湊值 $x$ 產製簽章值$(R_e, y_e)$。之後，可以使用公鑰 $P_e$ 根據公式(9)產製橢圓曲線點 $S_e$ 驗證簽章，當橢圓曲線點 $R_e$ 和橢圓曲線點 $S_e$ 相等時驗證簽章通過[10]。然而，Shor 提出可以在多項式時間複雜度內解決離散對數問題的量子演算法[6]，所以橢圓曲線密碼學方法也是易受量子攻擊密碼學方法之一。

$$P_e = p_e G \tag{6}$$
$$R_e = r_e G \tag{7}$$
$$y_e \equiv (x + p_e x_r)/r_e \ (\bmod \ n_e) \tag{8}$$
$$S_e = (x/y_e)G + (x_r/y_e)P_e \tag{9}$$

## 3. 後量子密碼學—基於格的密碼學

基於格的密碼學方法是後量子密碼學的主流方法之一。其中，美國國家標準暨技術研究院已經將 Kyber、Dilithium 和 Falcon 等多個基於格的密碼學方法在後量子密碼學標準化進程第三輪評選為標準方法[7]。為說明基於格的密碼學方法，本小節以 NTRU 演算法作為例。首先，選擇兩個整數 $b_1$ 和 $b_2$，並且使其最大公因數為 1。然後，隨機生成兩個方程式 $f_1$ 和 $f_2$，確保 $f_1$ 對於 $b_1$ 和 $b_2$ 都存在乘法逆元，如公式(10)和公式(11)，以產製私鑰$\{f_1, F_{b_1}, F_{b_2}\}$。之後，使用方程式 $F_{b_2}$ 和 $f_2$ 根據公式(12)生成公鑰 $P_n$ [11]。在加密過程中，將明文 $x$ 編碼為函數 $f_x$，並產製隨機函數 $f_r$，根據公式(13)產製密文 $f_y$。在解密過程中，可以使用私鑰$\{f_1, F_{b_1}, F_{b_2}\}$解密密文 $f_y$，並根據公式(14)獲得函數 $f_x$ [12]。

$$F_{b_1} \times f_1 \equiv 1 (\bmod \ b_1) \tag{10}$$
$$F_{b_2} \times f_1 \equiv 1 (\bmod \ b_2) \tag{11}$$
$$P_n \equiv F_{b_2} \times f_2 (\bmod \ b_2) \tag{12}$$
$$f_y \equiv b_1 f_r P_n + f_x (\bmod \ b_2) \tag{13}$$
$$f_x \equiv F_{b_1} [f_1 f_y (\bmod \ b_2)] (\bmod \ b_1) \tag{14}$$

## 4. 後量子密碼學—基於雜湊的密碼學

基於雜湊的密碼學方法也是後量子密碼學的另一個主流方法。其中，SHPINCS+已被美國國家標準暨技術研究院在後量子密碼學標準化進程第三輪評選為標準方法之一[7]。本小節以 Winternitz One-Time Signature (WOTS)演算法[13]為例，說明基於雜湊的密碼學方法。首先，選擇雜湊函數 $h$，例如：安全雜湊算法(Secure Hash Algorithm-256, SHA-256)。設置參數 $c$ 限制傳

輸訊息的最大值。產製隨機數 $p_w$ 作為私鑰，通過對 $p_w$ 執行 $c$ 次雜湊計算來產製公鑰 $P_w$，如公式(15)所示。在產製簽章過程中，對明文雜湊值 $x$ 產製簽章，對 $p_w$ 執行 $x$ 次雜湊計算產製簽章值 $y_w$，如公式(16)所示。在驗證簽章過程中，採用公式(17)對簽章值 $y_w$ 執行 $(c-x)$ 次雜湊計算；當公式(17)的結果等於公鑰 $P_w$ 時，驗證簽章通過[14]。

$$P_w = h^c(p_w) \tag{15}$$

$$y_w = h^x(p_w) \tag{16}$$

$$h^{c-x}(y_w) \tag{17}$$

## 5. 後量子密碼學—基於編碼的密碼學

在後量子密碼學標準化進程第四輪評選中，美國國家標準暨技術研究院考慮包括 BIKE、Classic McEliece、HQC、以及 SIKE 等多個基於編碼的密碼學方法。為說明基於編碼的密碼學，本小節以 McEliece 密碼學方法[15]為例。McEliece 密碼學方法產製三個矩陣：一個混淆器(Scrambler) $M_1$、一個生成矩陣(Generator Matrix) $M_2$、以及一個置換矩陣(Permutation Matrix) $M_3$。這些矩陣，表示為 $\{M_1, M_2, M_3\}$ 作為私鑰，而這些矩陣的乘積 $P_m$ 則作為公鑰(公式(18)所示)[16]。在加密過程中，將明文 $x$ 編碼為矩陣 $M_x$，並產製隨機矩陣 $M_r$，根據公式(19)生成密文 $M_y$。在解密過程中，私鑰的逆矩陣被計算和表示為 $\{M_1^{-1}, M_4, M^{-1}\}$，其中 $M_4$ 是基於 $M_2$ 的解碼矩陣。此外，基於 $M_2$ 產製錯誤校正函數 $f_c$，用於去除隨機矩陣 $M_r$。因此，可以採用公式(20)解密密文 $M_y$，獲得明文 $M_x$ [17]。

$$P_m = M_1 M_2 M_3 \tag{18}$$

$$M_y = M_x P_m + M_r \tag{19}$$

$$M_x = M_1^{-1} M_4 \left[ f_c \left( M_y M_3^{-1} \right) \right] \tag{20}$$

## 6. 後量子密碼學—基於多變量的密碼學

基於多變量的密碼學方法是傑出的後量子密碼學方法之一。例如，Matsumoto-Imai (MI)密碼系統是基於多變量的密碼學領域中著名的方法，可以基於函數和反函數的操作進行加密/解密[18]和產製簽章/驗證簽章[19]。在產製金鑰過程中，MI 密碼系統涉及創建兩個具有 $L$ 個元素的單變量$(k–1)$次方程式：$F_1$ 和 $F_3$，以及一個單變量 $k$ 次方程式 $G$。$F_1$ 和 $F_3$ 必須都具有反函數，分別表示為 $F_1^{-1}$ 和 $F_3^{-1}$。對於函數 $F_3$，其輸入是一個單變量$(k–1)$次方程式 $q_x$。設置參數 $t_1$，滿足 $1 < t_1 < k$ 的特性。基於 $t_1$，建立一個非線性函數 $F_2$，計算為 $F_3(q_x)$ 的 $2^{t_1}+1$ 次方，如公式(21)所示。其中，函數 $F_1$、$F_2$、$F_3$ 和 $G$ 被採用為私鑰。對於產製簽章，明文雜湊值 $x$ 編碼為函數 $q_x$，並作函數 $F_3$ 的輸入；之後，簽章 $q_y$ 可以通過公式(22)執行。此外，$F_2$ 存在反函數 $F_4$ 用於產製公鑰，由公式(23)表示。對於驗證簽章，可以採用公鑰 $F_3^{-1}\left(F_4\left(F_1^{-1}(q_y)\right)\right)$ 使用公式(24)來驗證簽章 $q_y$。當公式(24)的結果等於 $q_x$ 時，驗證簽章通過[19]。

$$F_2 = \left(F_3(q_x)\right)^{L^{t_1}+1} (\bmod\ G) \tag{21}$$

$$q_y = F_1\left(F_2(F_3(q_x))\right) \tag{22}$$

$$F_4 = \left(F_1^{-1}(q_y)\right)^{t_2} \pmod{G}, \text{where } (L^{t_1} + 1)t_2 \equiv 1 \left(\bmod \left(L^k - 1\right)\right) \tag{23}$$

$$q_x = F_3^{-1}\left(F_4\left(F_1^{-1}(q_y)\right)\right) \tag{24}$$

# 三、基於後量子密碼學的 X.509 資安憑證

第1節介紹 X.509資安憑證，並且討論如何將後量子密碼學演算法結合到 X.509憑證中。第2節說明每個後量子密碼學演算法及其參數，描述物件識別碼。最後，在第3節中提供基於後量子密碼學的 X.509資安憑證的案例研究。

## 1. X.509 資安憑證格式

X.509 資安憑證格式包括版本號、序號、簽發者、主體、有效期自、有效期到、公開金鑰、數位簽章演算法、簽章和指紋。在主體內，可以設定詳細資訊，如：通用名稱、組織名稱、部門名稱、地點等[20]。

簽發者表示簽發該憑證的實體，其中根憑證中心是自簽的，而對於終端設備的憑證是由憑證中心發行的。簽署的內容主要包含憑證持有人的相關信息、有效期自、有效期到的憑證效期。公開金鑰欄位包含憑證持有人的公開金鑰，可以用來驗證由憑證持有人簽章訊息。

簽章演算法欄位顯示憑證中心簽署憑證的數位簽章演算法。為了產生基於後量子密碼學的 X.509 資安憑證，可以在簽章演算法欄位中指定後量子密碼學演算法。簽章欄位顯示憑證中心對憑證的簽章，可用於驗證憑證的真實性。最後，指紋欄位用於驗證憑證的完整性。

## 2. 物件識別碼和數位簽章演算法

在簽章演算法欄位裡會具體指出的數位簽章演算法及其相關參數(如：RSA with SHA256、ECDSA with P-256 and SHA256 等)。為了達到標準化，為每個演算法及其相關參數各別定義一個物件識別碼。表 1 列出常用的物件識別碼[21]-[23]，並且本研究採用已編制用於後量子密碼學演算法的物件識別碼資訊。這些物件識別碼可以用於實作 X.509 憑證的簽章演算法欄位。由於目前 NIST 標準主要包含基於格的數位簽章演算法(如：Dilithium 和 Falcon)和基於雜湊的數位簽章演算法(如：SPHINCS+)，所以本研究主要實作和比較已成為標準的後量子密碼學演算法。

## 3. 案例研究

本節提供案例研究，說明基於後量子密碼學的 X.509 資安憑證。在此案例中，存在一個自簽其憑證的根憑證中心，以及一個由根憑證中心簽發的終端設備憑證。詳述如下。

根憑證中心的 X.509 憑證如圖 1 所示。在簽章演算法欄位中，使用物件識別碼是 1.3.9999.3.1 (即數位簽章演算法是 Falcon 且參數為 512)。此外，終端設備的 X.509 憑證如圖 2 所示。在簽章演算法欄位中，可以找到物件識別碼是 1.3.9999.3.1(即數位簽章演算法是 Falcon 且參數為 512)。

表 1 物件識別碼和數位簽章演算法及其相關參數[21]-[23]

| 物件識別碼 | 數位簽章演算法 | 參數 | 抵抗量子計算攻擊的安全等級 |
|---|---|---|---|
| 1.2.840.113549.1.1.11 | RSA | 2048 and SHA256 | 不安全 |
| 1.2.840.113549.1.1.12 | RSA | 3072 and SHA384 | 不安全 |
| 1.2.840.113549.1.1.13 | RSA | 4096 and SHA512 | 不安全 |
| 1.2.840.10045.4.3.2 | ECDSA | P-256 and SHA256 | 不安全 |
| 1.2.840.10045.4.3.3 | ECDSA | P-384 and SHA384 | 不安全 |
| 1.2.840.10045.4.3.4 | ECDSA | P-521 and SHA512 | 不安全 |
| 1.3.9999.3.1 | Falcon | 512 | 1 |
| 1.3.9999.3.4 | Falcon | 1024 | 5 |
| 1.3.6.1.4.1.2.267.7.4.4 | Dilithium | 2 | 2 |
| 1.3.6.1.4.1.2.267.7.6.5 | Dilithium | 3 | 3 |
| 1.3.6.1.4.1.2.267.7.8.7 | Dilithium | 5 | 5 |
| 1.3.9999.6.4.4 | SPHINCS+ | shake_128f | 1 |
| 1.3.9999.6.5.3 | SPHINCS+ | shake_192f | 3 |
| 1.3.9999.6.6.3 | SPHINCS+ | shake_256f | 5 |

# 四、實驗比較與討論

為驗證基於後量子密碼學的 X.509資安憑證相關效能，本研究採用各種標準後量子密碼學演算法來實作 X.509憑證，並比較憑證長度、產製憑證時間、以及產製簽章和驗證簽章時間。第1節描述實驗環境，第2節~第4節描述比較結果。

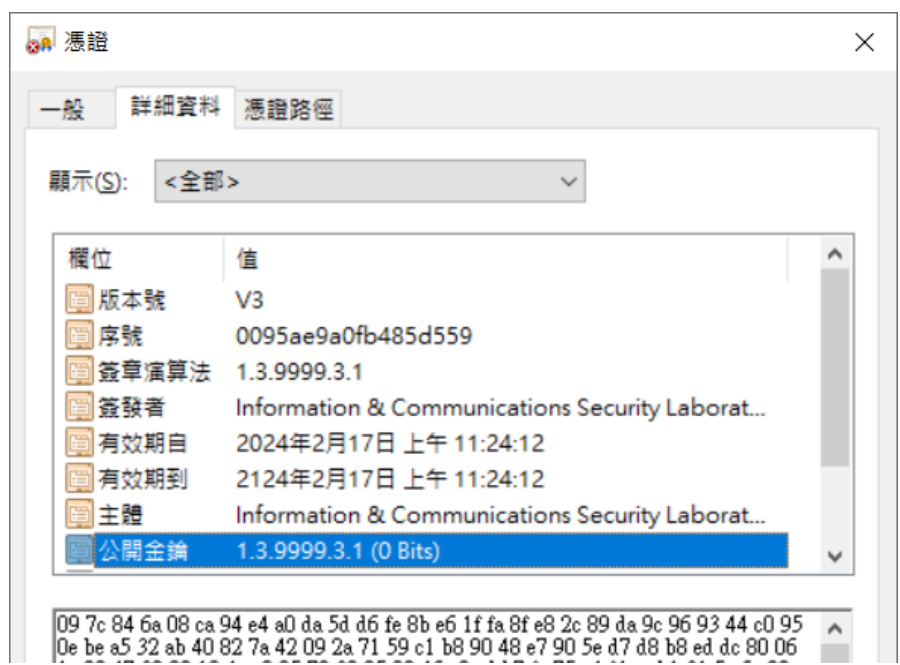

圖 1 根憑證中心的 X.509憑證

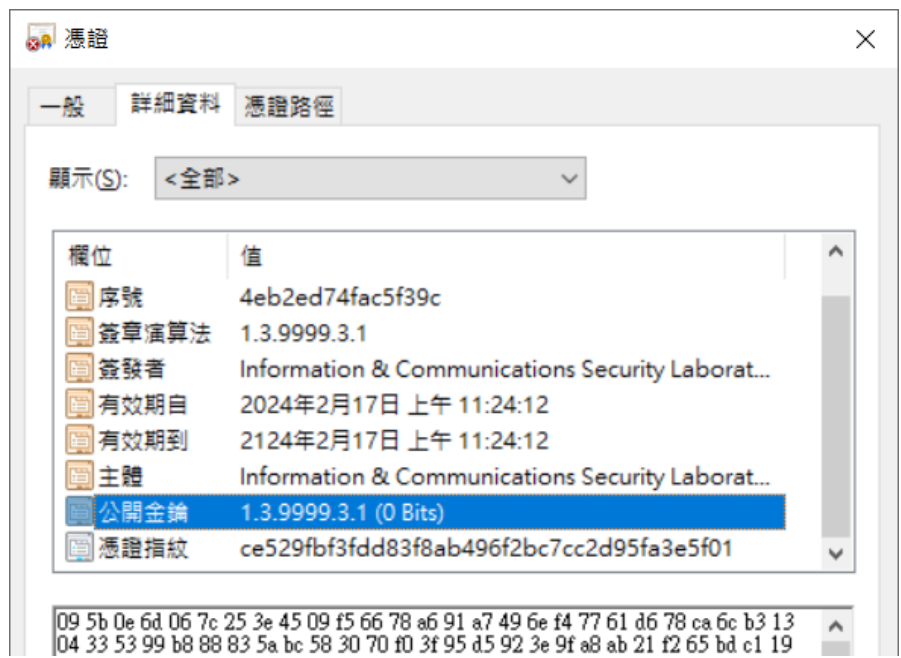

圖 2 終端設備的 X.509憑證

1.  實驗環境

在實驗環境中，按照表 1 所列的演算法及其參數進行實作。筆記型電腦的硬體和軟體資訊如下：CPU 是 Intel(R) Core(TM) i7-10510U、RAM 是 8 GB、OpenJDK 版本是 18.0.2.1、以及 JAR 函式數是 Bouncy Castle Release 1.72。並且採用 NIST 定義的安全等級[24]來描述安全性。

2.  憑證長度比較

表 2 實作各種數位簽章演算法，並且比較結合各種數位簽章演算法時，根憑證中心 X.509 憑證的長度。由實驗結果可以觀察到由於橢圓曲線數位簽章演算法有較短的公鑰和簽章，所以憑證長度最短。然而，由於橢圓曲線數位簽章演算法容易受到量子計算攻擊，所以橢圓曲線數位簽章演算法不安全。

在後量子密碼學系列的數位簽章演算法中，由於基於格的數位簽章演算法 Falcon 具有較短的簽章，所以可以產生較短的憑證長度。另外，實驗結果指出基於雜湊的數位簽章演算法 SPHINCS+產製的憑證長度最長。

3.  產製憑證時間比較

圖 3 實作各種數位簽章演算法，並且比較結合各種數位簽章演算法的產製憑證時間。本節主要以產製根憑證中心 X.509 憑證進行評估。產製憑證的過程主要包括產製金鑰對和產製簽章。研究結果顯示，RSA 和基於雜湊的數位簽章演算法 SPHINCS+產製憑證需要較長的計算時間。雖然橢圓曲線數位簽章演算法可以在較短的計算時間內產製憑證，但不具備抵抗量子計算攻擊的能力。在後量子密碼學系列的數位簽章演算法中，基於格的數位簽章演算法 Falcon 和 Dilithium 在產製憑證方面表現出優越效能。

表 2 根憑證中心憑證長度比較

| 物件識別碼 | 數位簽章演算法 | 參數 | 憑證長度(單位：KB) |
|---|---|---|---|
| 1.2.840.113549.1.1.11 | RSA | 2048 and SHA256 | 0.959 |
| 1.2.840.113549.1.1.12 | RSA | 3072 and SHA384 | 1.18 |
| 1.2.840.113549.1.1.13 | RSA | 4096 and SHA512 | 1.43 |
| 1.2.840.10045.4.3.2 | ECDSA | P-256 and SHA256 | 0.563 |
| 1.2.840.10045.4.3.3 | ECDSA | P-384 and SHA384 | 0.625 |
| 1.2.840.10045.4.3.4 | ECDSA | P-521 and SHA512 | 0.698 |
| 1.3.9999.3.1 | Falcon | 512 | 1.92 |
| 1.3.9999.3.4 | Falcon | 1024 | 3.39 |
| 1.3.6.1.4.1.2.267.7.4.4 | Dilithium | 2 | 4.06 |
| 1.3.6.1.4.1.2.267.7.6.5 | Dilithium | 3 | 5.54 |
| 1.3.6.1.4.1.2.267.7.8.7 | Dilithium | 5 | 7.44 |
| 1.3.9999.6.4.4 | SPHINCS+ | shake_128f | 17.1 |
| 1.3.9999.6.5.3 | SPHINCS+ | shake_192f | 35.2 |
| 1.3.9999.6.6.3 | SPHINCS+ | shake_256f | 49.1 |

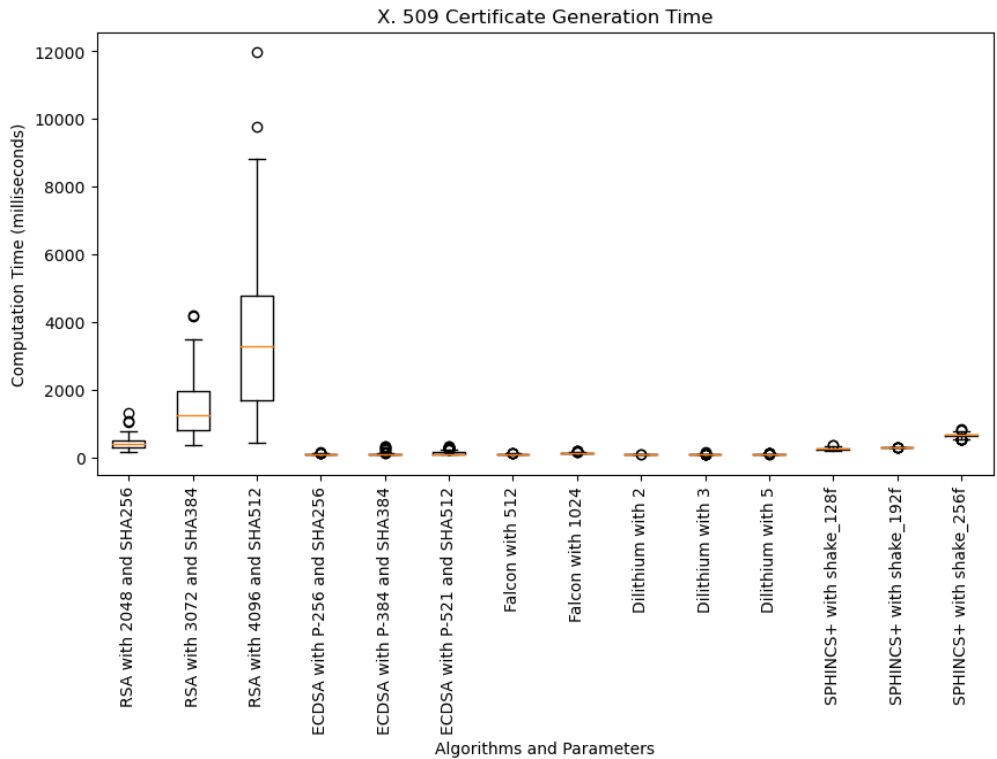

圖 3 根憑證中心 X.509憑證產製憑證時間比較

## 4. 產製簽章和驗證簽章時間比較

圖 4 實作各種數位簽章演算法，並且比較結合各種數位簽章演算法的產製簽章時間。雖然

RSA 和橢圓曲線數位簽章演算法在產製簽章方面表現較佳，但它們容易受到量子計算攻擊的影響。在後量子密碼學系列的數位簽章演算法中，雖然基於雜湊的數位簽章演算法 SPHINCS+具有抵抗量子計算攻擊的能力，但實驗結果顯示它在產製簽章方面需要更長的計算時間。因此，基於格的數位簽章演算法 Falcon 和 Dilithium 為更適合的解決方案，能夠擁有較短的產製簽章時間，同時可以抵禦量子計算攻擊。

圖 5 實作各種數位簽章演算法，並且比較結合各種數位簽章演算法的驗證簽章時間。雖然 RSA 在驗證簽章方面表現良好，但不具備抵抗量子計算攻擊的能力。另外，研究結果顯示，基於格的數位簽章演算法 Falcon 和 Dilithium 在驗證簽章方面表現較佳，可以提供比橢圓曲線數位簽章演算法提供更高的驗證簽章效率。此外，基於雜湊的數位簽章演算法 SPHINCS+的驗證簽章效率較差，因此可以考慮作為一個可行的替代方案。

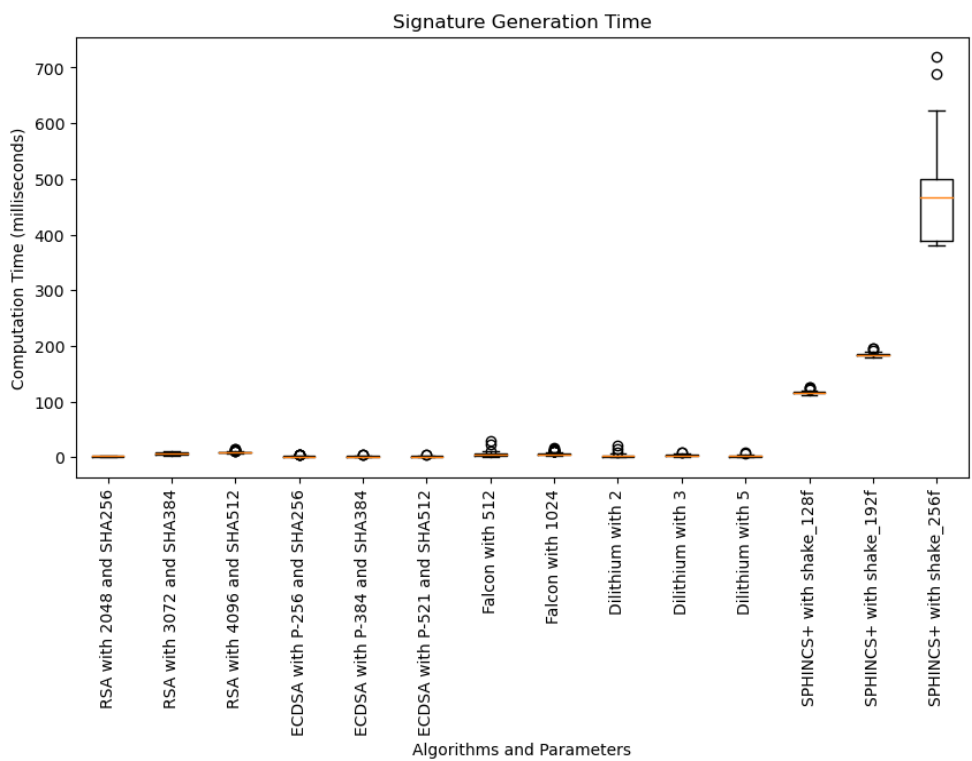

圖 4 產製簽章時間比較

## 五、結論與未來研究

由於現行主流的非對稱密碼學演算法(如：RSA 和橢圓曲線密碼學)在對抗量子計算攻擊方面存在不足，本研究探索結合後量子密碼學到 X.509資安憑證中的可行性。

本研究的主要貢獻概述如下：
(1). 探索並實作結合後量子密碼學到 X.509資安憑證中的解決方案。
(2). 比較後量子密碼學演算法和現行主流的非對稱密碼學演算法在產製 X.509憑證的效率。
(3). 比較後量子密碼學演算法和現行主流的非對稱密碼學演算法在產製簽章和驗證簽章的效率。

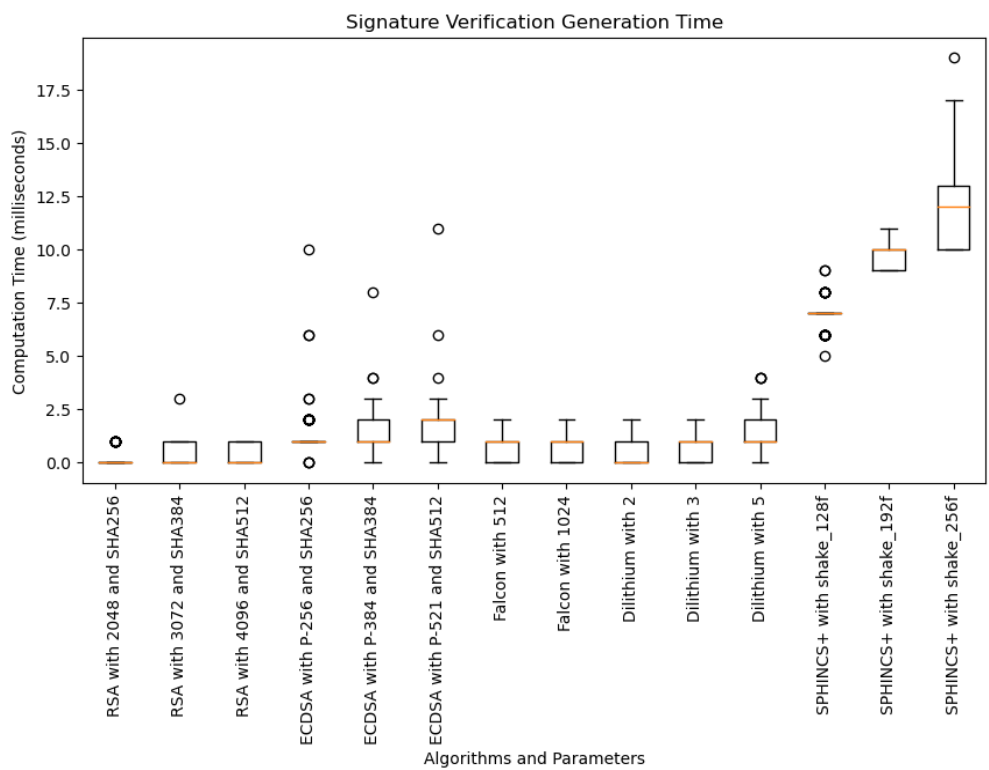

圖 5 驗證簽章時間比較

對於未來的研究，可以考慮將本研究提出的解決方案應用到傳輸層安全性協定(Transport Layer Security, TLS)等協議中，並在更廣泛在各個線上系統上進行效能驗證。

# 參考文獻